\documentclass[showpacs,amsmath,amssymb,twocolumn,prl,superscriptaddress]{revtex4-1}

\usepackage[dvips]{graphicx} 
\usepackage[colorlinks,linkcolor=blue,anchorcolor=blue,citecolor=blue]{hyperref}
\usepackage{amsfonts}
\usepackage{amssymb}
\usepackage{amscd}
\usepackage{amsmath}    
\usepackage{enumerate}
\usepackage{epsfig}
\usepackage{subfigure}
\usepackage{xcolor}
\usepackage{amsthm}
\usepackage{framed}

\usepackage{tcolorbox}
\usepackage{siunitx}
\usepackage{xspace}
\usepackage{array}
\usepackage{multirow}

\newcommand{\bra}[1]{\mbox{$\left\langle #1 \right|$}}
\newcommand{\ket}[1]{\mbox{$\left| #1 \right\rangle$}}

\newcommand{\comments}[1]{}
\newcommand{\CZ}{\textrm{CZ}}
\newcommand{\Tr}{\textrm{Tr}}

\begin{document}
	\title{
		Robust Self-Testing of Multiparticle Entanglement}
	
	\date{\today}
	
	\author{Dian Wu}
	
	\affiliation{Hefei National Laboratory for Physical Sciences at Microscale and Department of Modern Physics, University of Science and Technology of China, Hefei, Anhui 230026, China}
	\affiliation{CAS Center for Excellence and Synergetic Innovation Center in Quantum Information and Quantum Physics, University of Science and Technology of China, Shanghai 201315, China}
	
	\author{Qi Zhao}
	
	\affiliation{Joint Center for Quantum Information and Computer Science, University of Maryland, College Park, Maryland 20742, USA}
	
	\author{Xue-Mei Gu} 
	
	\author{Han-Sen Zhong}
	\affiliation{Hefei National Laboratory for Physical Sciences at Microscale and Department of Modern Physics, University of Science and Technology of China, Hefei, Anhui 230026, China}
	\affiliation{CAS Center for Excellence and Synergetic Innovation Center in Quantum Information and Quantum Physics, University of Science and Technology of China, Shanghai 201315, China}
	\author{You Zhou}
	\affiliation{School of Physical and Mathematical Sciences, Nanyang Technological University, Singapore 637371, Republic of Singapore}
	\author{Li-Chao Peng}
	\author{Jian Qin}
	\author{Yi-Han Luo}
	\author{Kai Chen}
	\author{Li Li}
	\author{Nai-Le Liu}
	\author{Chao-Yang Lu}
	\author{Jian-Wei Pan}
	
	\affiliation{Hefei National Laboratory for Physical Sciences at Microscale and Department of Modern Physics, University of Science and Technology of China, Hefei, Anhui 230026, China}
	\affiliation{CAS Center for Excellence and Synergetic Innovation Center in Quantum Information and Quantum Physics, University of Science and Technology of China, Shanghai 201315, China}

	\begin{abstract}

		Quantum self-testing is a device-independent way to certify quantum states and measurements using only the input-output statistics, with minimal assumptions about the quantum devices. Because of the high demand on tolerable noise, however, experimental self-testing was limited to two-photon	systems.
		Here, we demonstrate the first robust self-testing for multi-photon genuinely entangled quantum states. We prepare two examples of four-photon graph states, the Greenberger-Horne-Zeilinger (GHZ) states with a fidelity of 0.957(2) and the linear cluster states with a fidelity of 0.945(2). Based on the observed input-output statistics, we certify the genuine four-photon entanglement and further estimate their qualities with respect to realistic noise in a device-independent manner.
	\end{abstract}

	\maketitle
	

	Multipartite quantum entanglement that exhibits correlations without a classically analog \cite{einstein1935can, bell1964einstein} plays a prominent role in understanding the quantum foundations as well as for quantum technologies. One of the canonical examples is the graph state \cite{Hein2004Multiparty} that comprises many popular states including Greenberger-Horne-Zeilinger (GHZ) states \cite{ greenberger1990bell} and cluster states \cite{Raussendorf2001One}. These genuine multipartite entangled states act as incredibly useful resources for quantum information applications such as quantum computation \cite{Raussendorf2001One}, error correction \cite{schlingemann2001quantum}, quantum cryptography \cite{markham2008graph}, and quantum metrology \cite{shettell2020graph}. However, certifying that a given multipartite quantum system is indeed working as intended is very complicated and challenging, which is a crucial step for future application implementations.

	Self-testing that was originated in device-independent (DI) scenarios and named by Mayers and Yao \cite{Mayers98, mayers2004self}, represents the strongest possible form of verification of quantum systems \cite{vsupic2020self}. Different from certification techniques such as tomography \cite{Vogel1989Determination,Paris2004esimation} and entanglement witnesses \cite{jungnitsch2011entanglement, Friis2019Reviews}  
	which require the characterization or assumption for the devices, self-testing allows us to verify the underlying functioning of a given quantum apparatus in a black-box fashion 
	without the need to know its inner workings. Specifically, it relies only on the observed classical input-output correlation statistics that maximally violate Bell-like inequalities \cite{Brunner2014nonlocality}. 
	A well-known example is that the maximal violation of the Clauser-Horne-Shimony-Holt (CHSH) inequality  \cite{clauser1969proposed} uniquely implies (self-tests) the presence of a two-qubit maximally entangled state \cite{bardyn2009device}. Such DI certifications have been widely studied for self-testing different quantum states and measurements with distinct Bell inequalities in many scenarios (please see Ref. \cite{vsupic2020self} for a review). 
	From a practical perspective, due to the unavoidable noise and finite date effect, one could not exactly achieve the  maximal violation of Bell inequalities. Importantly, when the violations are only close  to the ideal ones, the underlying states and measurements should also be close to the desired ones. This is referred to as robust self-testing \cite{mckague2012robust}. To be more specific, one can still infer the underlying states and estimate the bound on the fidelity of the actual state with respect to the target one even when the maximal violation of Bell inequalities is only approximately met. 
	
	Considerable theoretical efforts have been made towards constructing different Bell inequalities for self-testing protocols and improving the robustness bounds for various quantum states \cite{vsupic2020self, yang2014robust, wu2014robust, kaniewski2016analytic, kaniewski2017self, coladangelo2017all, breiner2018parallel, baccari2018scalable, zhao2020constructing}. To self-test any multipartite graph states with a large number of particles, protocols require an efficient scaling in terms of the computational resources are crucial for experimental implementations \cite{vsupic2020self}. Remarkably, a new family of Bell inequalities that are both scalable and robust \cite{baccari2018scalable,zhao2020constructing} with a linear number of correlations improved the robustness self-testing fidelity bounds for multipartite graph states, which presents a significant reduction of experimental efforts needed.

	Despite of many Bell nonlocality demonstrations \cite{Brunner2014nonlocality}, the self-testing relevant experimental demonstration is little to know. 
	This is because robust self-testing generally requires nearly maximal violations which depends on the nearly perfect state and  preparations, whereas Bell nonlocality demonstrations only require the existence of violations.
	To date, only a few optical experiments have been implemented principally to self-test quantum states such as a Bell state distributed over 398 meters in a fully DI manner \cite{bancal2021self}, partially entangled pairs of qubits \cite{zhang2019experimental, gomez2019experimental, goh2019experimental}, bipartite and tripartite qubit states \cite{PhysRevLett.121.240402, li2019experimental}, two-qutrit entangled states \cite{wang2018multidimensional}, and two copies of bipartite states \cite{agresti2020experimental}. 
	The robust self-testing of quantum states is only experimentally studied in two-photon or the copies of two-photon entangled systems  \cite{PhysRevLett.121.240402,zhang2019experimental,agresti2020experimental}.
	However, to the best of our knowledge, experimental robust self-testing for genuinely multipartite entangled graph states beyond two photons, especially the cluster states, remains unexplored.

	In this Letter, we perform a proof-of-concept demonstration of robust self-testing for two important examples of multipartite graph states -- a four-photon GHZ state and a four-photon linear cluster state, for the first time. We first prepare high quality four-photon entanglement sources as well as their mixture with various degrees of noise, and then perform measurement settings required by each local party (up to local isometries) \cite{baccari2018scalable,zhao2020constructing}. From the experimentally observed input-output statistics that violate certain Bell inequalities \cite{baccari2018scalable,zhao2020constructing}, we certify the genuine four-photon entanglement and infer their fidelities with respect to realistic noise according to the robustness bound in a DI manner. Our work promotes self-testing as a practical tool to develop quantum techniques for future applications such as quantum cryptography \cite{Mayers98,acin2007device} and delegated blind quantum computing \cite{reichardt2013classical,PhysRevLett.119.050503}.

	\textit{Theoretical proposal.--- }The robust self-testing strategy is based on a new family of Bell inequalities that can be maximally violated by any multipartite graph states \cite{baccari2018scalable,zhao2020constructing}. The new constructed Bell inequalities are based on stabilizers. 
	\begin{figure}[!t]
		\centering
		\includegraphics[scale=0.24]{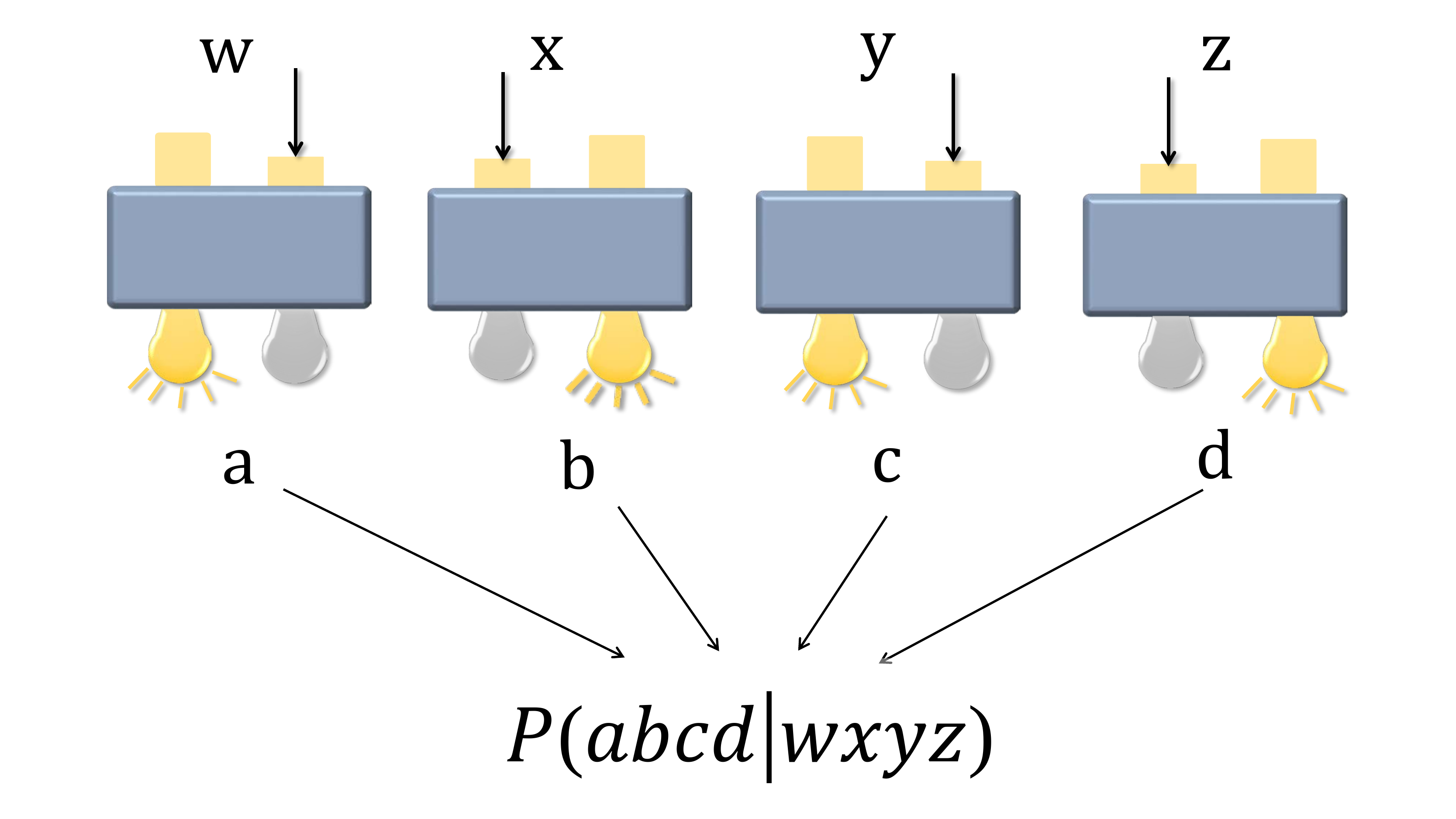}
		\caption{Four-party Bell inequalities: 
			In each round of Bell inequalities, each party receives a classical bit $w, x, y, z$ and output a classical bit $a, b, c, d$. 
			A score is obtained according to the conditional probabilities $P(abcd|wxyz)$ and a predetermined rule.}\label{fig:concept}
	\end{figure}
	Now let us first recall the definitions of graph states, the structure of which can be described in a concise and fruitful way by mathematical graphs \cite{Hein2004Multiparty}. Given a graph $\mathcal{G}=(V,E)$ with the vertex set $V=\{1,2,\dots,N\}$ and the edge set $E\subseteq[V]^2$, each vertex in $V$ represents a qubit prepared in the state $\ket{+}=(\ket{0}+\ket{1})/\sqrt{2}$ and each edge in $E$ stands for a controlled-Z gate (i.e., $\CZ_{\{i,j\}}=\ket{0}\bra{0}_i\otimes \mathbb{I}_j+\ket{1}\bra{1}_i\otimes Z_j$ and $Z_j$ is Pauli Z operator for qubit $j$) applied between the two connected qubits, yielding a graph state $\ket{\psi_\mathcal{G}}=\prod_{(i,j)\in E}\CZ_{\{i,j\}}\ket{+}^{\otimes N}$. Equivalently, graph states can be uniquely determined by $N$ generators, i.e., $G_i=X_i\bigotimes_{j\in n_i} Z_j$ ($X_j$ is Pauli X operator for qubit $i$) with $n_i$ being the neighbor set of vertex $i$.

	All the stabilizers can be generated by the multiplication of these generators $S_k=\prod_i G_i$, which satisfy $S_k\ket{\psi_\mathcal{G}}= \ket{\psi_\mathcal{G}}$. The property of stabilizers can be utilized to verify the graph states and construct entanglement witness efficiently \cite{Toth2005Detecting, zhou2019detecting}. Assisted with the stabilizers of graph states, new families of Bell inequalities are introduced \cite{baccari2018scalable,zhao2020constructing}, which can be used for robust self-testing.  
	We now construct the Bell-like inequalities for four-qubit graph states with the following equations Eq.~\eqref{Eq:GHZ} and Eq.~\eqref{Eq:Cluster} ($\mathcal{B}_1$, $\mathcal{B}_2$, and $\mathcal{B}_3$ for GHZ states; $\mathcal{B}_4$, $\mathcal{B}_5$, and $\mathcal{B}_6$ for linear cluster states).
	\begin{widetext}
		\begin{equation}\label{Eq:GHZ}
		GHZ \left\{
		\begin{aligned}
		&\mathcal{B}_1: \big\langle(A_1+B_1)B_2B_3B_4 \big \rangle+  \big\langle(A_1-B_1)A_2 \big\rangle + \big\langle A_2A_3  \big\rangle+ \big\langle A_2A_4 \big \rangle
		\le \beta_{C,1}=4, \\
		& \mathcal{B}_2 : 2\big\langle(A_1+B_1)B_2B_3B_4 \big \rangle+  \big\langle(A_1-B_1)A_2 \big\rangle +
		\big\langle(A_1-B_1)A_3 \big\rangle+
		\big\langle A_2A_4 \big \rangle
		\le \beta_{C,2}=5, \\
		&   \mathcal{B}_3: 3\big\langle(A_1+B_1)B_2B_3B_4 \big \rangle+  \big\langle(A_1-B_1)A_2 \big\rangle +
		\big\langle(A_1-B_1)A_3 \big\rangle+
		\big\langle(A_1-B_1)A_4 \big\rangle
		\le \beta_{C,3}=6 .\\
		\end{aligned}
		\right.
		\end{equation}
		\begin{equation}\label{Eq:Cluster}
		Cluster \left\{
		\begin{aligned}
		&\mathcal{B}_4: \big  \langle(A_1+B_1)B_2\big  \rangle+ \big  \langle(A_1-B_1)A_2B_3  \big  \rangle +\big  \langle B_2A_3B_4 \big \rangle+\big  \langle B_3A_4 \big  \rangle \le \beta_{C,1}=4,\\
		&  \mathcal{B}_5: \big  \langle A_1(A_2-B_2)\big  \rangle+ 2\big  \langle B_1(A_2+B_2)B_3  \big  \rangle +\big  \langle (A_2-B_2)A_3B_4 \big \rangle+\big  \langle B_3A_4 \big  \rangle \le \beta_{C,2}=5,\\
		& \mathcal{B}_6: \big  \langle A_1(A_2-B_2)\big  \rangle+ \big  \langle B_1(A_2+B_2)B_3  \big  \rangle +\big  \langle (A_2-B_2)A_3B_4 \big \rangle+\big  \langle B_1(A_2+B_2)A_4 \big  \rangle \le \beta_{C,3}=4.
		\\
		\end{aligned}
		\right.
		\end{equation}
	\end{widetext}
	Here $\beta_{C,i}$ are the classical bounds for Bell inequalities. For the GHZ states, the optimal quantum bound  in Eq.~\eqref{Eq:GHZ} can be reached by taking $A_1=\frac{X+Z}{\sqrt{2}}$, $B_1=\frac{X-Z}{\sqrt{2}}$, and $A_i=X$, $B_i=Z$ when $i \ne 1$. The maximal quantum values $\beta_{Q,i}$ can reach $\beta_{Q,1}=2+2\sqrt{2}$, $\beta_{Q,2}=1+4\sqrt{2}$, $\beta_{Q,3}=6\sqrt{2}$. For the cluster states, the optimal quantum measurement settings are  $A_1=\frac{X+Z}{\sqrt{2}}$, $B_1=\frac{X-Z}{\sqrt{2}}$, and $A_i=X$, $B_i=Z$ when $i \ne 1$ for $\mathcal{B}_4$. For $\mathcal{B}_5$ and $\mathcal{B}_6$, the optimal settings are $A_2=\frac{X+Z}{\sqrt{2}}$, $B_2=\frac{X-Z}{\sqrt{2}}$, and $A_i=X$, $B_i=Z$ when $i \ne 2$. The quantum bounds are $\beta_{Q,4}=2+2\sqrt{2}$, $\beta_{Q,5}=1+4\sqrt{2}$, $\beta_{Q,6}=4\sqrt{2}$.

	We now consider the scenario in which four distant clients share a four-qubit graph state. As shown in Fig.~\ref{fig:concept}, for each experimental trial, the clients perform a choice of measurement settings upon receiving classical random inputs $w$, $x$, $y$ and $z$ (0 or 1) to produce classical outputs $a$, $b$, $c$ and $d$ ($+1$ or $-1$), respectively. After repeating the experiment sufficiently many times and collecting the observed statistics, the joint conditional probabilities $P(abcd|wxyz)$  can be estimated for the Bell inequalities in Eq.~\eqref{Eq:GHZ} and Eq.~\eqref{Eq:Cluster}, revealing the properties of the prepared states. The violation of classical bounds (i.e., $\langle \mathcal{B}_i\rangle>\beta_{C,i}$) implies the presence of entanglement and nonlocality in the test state $\rho$. 
	When the expected Bell value reaches the maximum quantum bound, $\langle \mathcal{B}_i \rangle =\beta_{Q,i}$, we can verify the prepared state $\rho$ is the target graph state $\ket{\psi_G}\bra{\psi_G}$, up to local isometries. 
	Meanwhile, one can also estimate the fidelity between underlying state  the measured state $\rho$ and the target graph state $\psi_G$ (under local isometry), on account of the Bell inequality value,
	\begin{equation}
	\begin{aligned}
	F=\max_{\Lambda=\Lambda_1\otimes\Lambda_2\cdots\Lambda_N } \bra{\psi_G} \Lambda(\rho) \ket{\psi_G},
	\end{aligned}
	\end{equation}
	where $\Lambda_i$ is the local channel on the $i$-th party. 
	Using the techniques in Refs.~\cite{kaniewski2016analytic,baccari2018scalable,zhao2020constructing},
	the fidelity can be lower bounded via
	\begin{equation}\label{Eq:fidelity bound}
	F\geq s_i\langle\mathcal{B}_i\rangle+\mu_i,
	\end{equation}
	where $\langle\mathcal{B}_i\rangle$ is the observed Bell value in  inequality $\mathcal{B}_i$. The coefficients $s_i$ and $\mu_i$ are obtained via optimizing all possible measurement angles (more details see Supplemental Material \cite{supp}). Importantly, $F>\frac{1}{2}$ indicates the underlying prepared state is genuinely entangled \cite{GUHNE2009detection}.

	\begin{figure}[!t]
		\centering
		\includegraphics[width=0.5\textwidth]{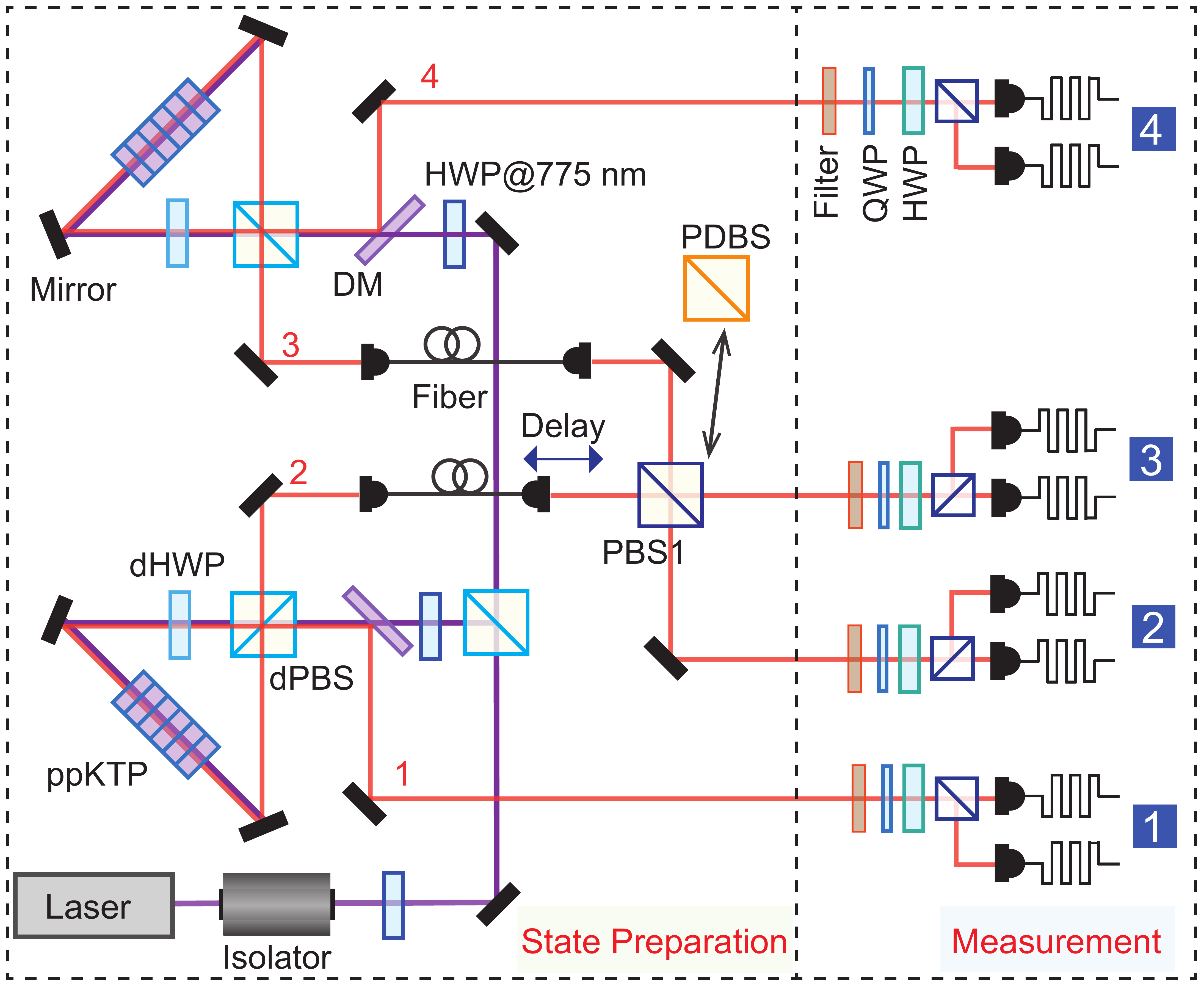}
		\caption{The experimental setup. A pulsed Ti-sapphire laser as the pump beam (repetition rate 80 MHz, center wavelength 775 nm, an average power $\sim0.9$ W, temporal duration $\sim130$ fs) passes through an optical isolator and half-wave plate (HWP, 775 nm) and then is split into two beams by a dual-wavelength polarizing beam splitter (dPBS, 775/1550 nm). Two periodically poled KTiOPO4 (ppKTP) crystals in polarization-based Sagnac Interferometers are pumped by the split pump beams for producing polarization-entangled photon pairs \cite{jin2014pulsed,PhysRevA.73.012316}. The remaining pump lights are removed by dichroic mirrors (DM). Fine adjustments of the delays between different paths are made so that the photons arrive at the polarizing beam splitter (PBS1) or polarization dependent beam splitter (PDBS) simultaneously. When there is only one photon in each path after PBS1 (or PDBS) and conditionally on the four-fold coincidences \cite{PhysRevLett.95.210502,RevModPhys.84.777}, we obtain a four-photon GHZ state (or linear cluster state). All the photons are spectrally filtered by filters and the outputs are detected by superconducting nanowire single photon detectors with average detector efficiency of $79\%$ where all the four-photon coincidences are recorded by a multi-channel coincidence count unit. QWP and dHWP represent quarter wave plate and dual-wavelength half-wave plate.}
		\label{fig:setup}
	\end{figure}
	
	\textit{Experimental realization.--- }As illustrated in Fig.~\ref{fig:setup}, with the pulsed laser pumping two ppKTP crystals in polarization-based Sagnac loops \cite{jin2014pulsed,PhysRevA.73.012316}, two entangled photon pairs in the form of $\ket{\Psi}=(\cos(\theta)\ket{00}+\sin(\theta)\ket{11})/\sqrt{2}$ are generated, where $\ket{0}$ and $\ket{1}$ represent horizontal $\ket{H}$ and vertical $\ket{V}$ polarization, respectively and $\theta$ is controlled by the pumping polarization \cite{Fedrizzi:07}. These two entangled pairs in spatial mode 2 and 3 are then superimposed at PBS1 (or PDBS). In order  to  ensure  the  photons  in  mode  2  and  3 arrive  at PBS1 (PDBS) simultaneously, we employ single-mode fibers and mount each fiber output end on a translation stage to precisely compensate time delay. When there is only one photon in each path after PBS1 (or PDBS) and conditionally on a four-fold coincidence detection, a four-photon entangled GHZ (or cluster) state is generated \cite{RevModPhys.84.777}. 
	
	For demonstrating robust self-testing of graph states, we prepare four-qubit graph states with various degrees of noise, 
	\begin{equation}
	\rho_{\text{exp}}= \frac{1}{1+p} \ket{\psi}\bra{\psi} +\frac{p}{1+p}\rho_{\text{noise}},
	\end{equation}
	where the pure state $\ket{\psi}$ is either GHZ states $\ket{G_{4}}_{1234}=(\ket{0000}+\ket{1111})/\sqrt{2}$ or linear cluster states $\ket{C_4}_{1234}=(\ket{0000}+\ket{0011}+\ket{1100}-\ket{1111})/2$, $\rho_{\text{noise}}=\ket{\Psi_{12}}\bra{\Psi_{12}}\otimes \ket{\Psi_{34}}\bra{\Psi_{34}}$ is the noise state given from four-fold coincidence detection under the condition of no interference between $\ket{\Psi_{12}}$ and $\ket{\Psi_{34}}$, and $p$ is the noisy proportion.

	\begin{figure*}[!t]
		\centering
		\includegraphics[width=1\textwidth]{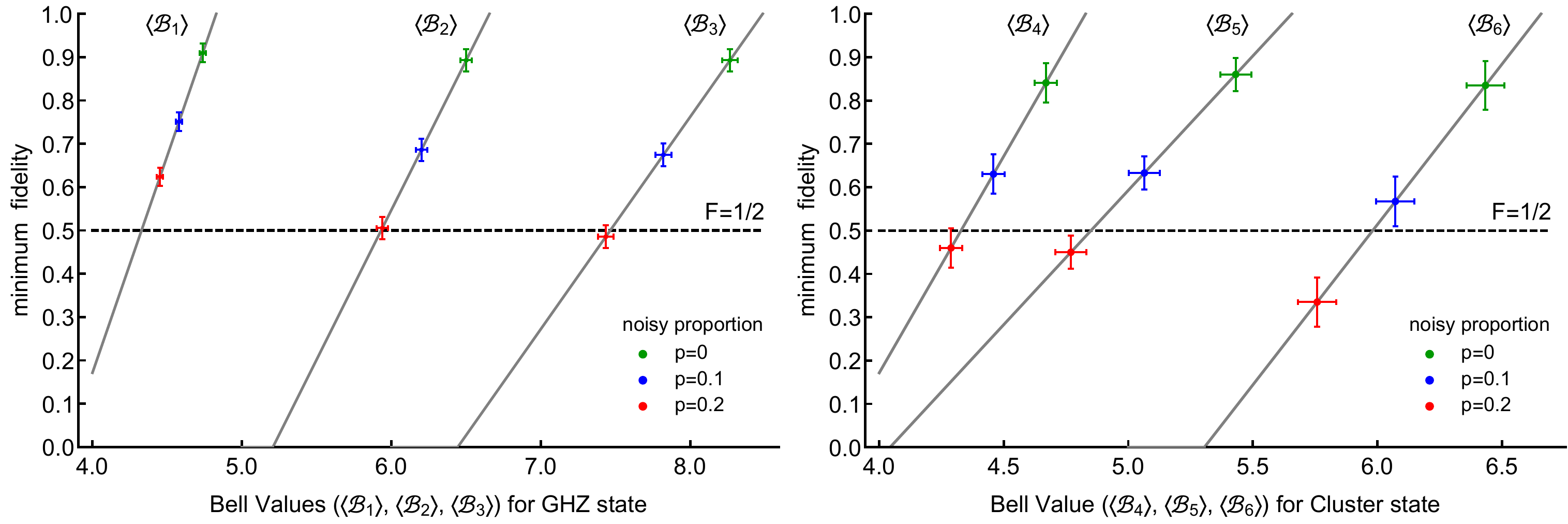}
		\caption{Minimum fidelity $F$ to the target graph state versus various Bell inequality values $\langle\mathcal{B}_1\rangle\sim\langle\mathcal{B}_6\rangle$. The dashed line $(F=1/2)$ is the boundary between the genuine entanglement $(F>1/2)$ and inconclusive $(F\le 1/2)$. The green, blue and red dots indicate different noisy proportion $p=0$, $p=0.1$ and $p=0.2$, respectively.} 
		Error bars indicate one standard deviation deduced from propagated Poissonian counting statistics of the raw detection events.\label{Fig:result}
	\end{figure*}
	
	To generate a four-photon GHZ state, two entangled pairs $\ket{\Psi_{12}}=\ket{\Psi_{34}}=(\ket{00}+\ket{11})/\sqrt{2}$ are required \cite{RevModPhys.84.777}. Then we estimated the fidelity of our prepared entangled GHZ state by entanglement witness \cite{PhysRevA.76.030305}, yielding $0.957(2)$ for the ideal situation with $p=0$ \cite{supp}. The GHZ state we produced is local-unitary equivalent to the states with same graph representations, where the local-unitary is $H_2H_3H_4$ ($H$ is a Hadamard gate, defined as $H =(X+Z)/\sqrt{2}$). Therefore, the quantum measurement settings for Bell inequalities in Eq.~\eqref{Eq:GHZ} are modified for the GHZ state $\ket{G_{4}}$ (please refer to Ref.~\cite{supp} ). The self-testing performance can be verified by the actual fidelity from the expected values $\big\langle\mathcal{B}_i \big \rangle$ of the modified Bell inequalities, as shown in Fig.~\ref{Fig:result}. We totally run $1.3\times 10^5$ trials for the modified Bell inequalities $\mathcal{B}_1$, $\mathcal{B}_2$ and $\mathcal{B}_3$ with different noisy probability $p$=0, 0.1, and 0.2, respectively. The corresponding results are also labeled in Fig.~\ref{Fig:result}. In the case of $p=0$, experimentally obtained Bell values are $\big\langle\mathcal{B}_1 \big \rangle =  4.74(2)$, $\big\langle\mathcal{B}_2 \big \rangle = 6.50(4)$, and $\big\langle\mathcal{B}_3 \big \rangle = 8.27(5)$ as well as their related fidelities which are $0.91(2)$, $0.89(3)$, and $0.89(3)$, respectively \cite{supp}. In our experiment, Bell inequality $\mathcal{B}_1$ shows the best fidelity estimation. Note that when $F>1/2$, we could also witness the genuine entanglement of prepared states. When $p=0$ and $0.1$, we can still witness the genuine entanglement via the three Bell inequalities in Eq.~\eqref{Eq:GHZ}. While in the case of $p=0.2$, only Bell inequality $\mathcal{B}_1$ allows us to witness the existence of genuine entanglement.

	To generate four-photon linear cluster state, two non-maximally entangled states $\ket{\Psi_{12}}=\ket{\Psi_{34}}=(\ket{00}+\sqrt{3}\ket{11})/2$ are required \cite{PhysRevLett.95.210502}. Instead of using PBS1 that transmits $\ket{H}$ and reflects $\ket{V}$ polarization for the generation of GHZ states, PDBS is employed for cluster states, where its transmission and reflection efficiencies for $\ket{V}$ and $\ket{H}$ photons are set to be $T_V=1/3$ and $R_V=2/3$, and $T_H=1$ and $R_H=0$, respectively. Also, the transmission rate of the $\ket{V}$ photons can be finely adjusted by tuning the photon incident angle on the PDBS. To ensure that photons in paths 2 and 3 arrive at the PDBS simultaneously, we perform a Hong–Ou–Mandel interferometer of two $\ket{V}$ polarized photons \cite{PhysRevLett.59.2044}. By measuring expectation values of 16 stabilizers \cite{GUHNE2009detection}, we estimate the fidelity of the linear cluster state $\ket{C_4}$ as $0.945(2)$ when $p=0$. In the demonstration of Bell inequalities $\mathcal{B}_4$, $\mathcal{B}_5$ and $\mathcal{B}_6$, we perform the quantum measurement settings via local-unitary $H_1H_4$. We then totally perform $3.1\times 10^4$ trials for the modified Bell inequalities. In the case of $p=0$, we experimentally obtain Bell values $\big\langle\mathcal{B}_4 \big \rangle=4.66(4)$, $\big\langle\mathcal{B}_5 \big \rangle=6.43(7)$ and  $\big\langle\mathcal{B}_6 \big \rangle=5.43(6)$ as well as the corresponding fidelities that are $0.84(4)$, $0.84(6)$, and $0.86(4)$, respectively \cite{supp}. In Fig.~\ref{Fig:result}, Bell inequality $\mathcal{B}_6$ shows the best performance for fidelity estimation. When $p=0$ and $0.1$, we can certify the genuine entanglement with the Bell inequalities in Eq.~\eqref{Eq:Cluster}. When $p=0.2$, we could only observe the existence of entanglement. From Fig.~\ref{Fig:result}, we see that the minimum fidelity curve varies with the Bell inequalities, it would be interesting to further improve the robust self-testing performance by proving a tighter minimum fidelity curve or proposing a more suitable Bell inequality tailored for the specific prepared states.

	We have for the first time demonstrated a robust self-testing for multi-qubit graph states with two important examples -- a four-photon GHZ state and a four-photon linear cluster state, based on a systematical framework where the constructed Bell inequalities are directly from stabilizers \cite{baccari2018scalable,zhao2020constructing}. 
	By preparing the high-quality four-photon entanglement source, we certify the genuine entanglement and estimate the fidelity at least $0.91(2)$ and $0.86(3)$ for GHZ state and cluster state, respectively, as well as their mixtures with various noise.
	We shall note that, our experiment is a proof of principle studies of the proposed scalable and robust self-testing for any multipartite graph states with the fair sampling assumption. 
	How to close the loopholes from multipartite Bell inequality in self-testing, such as detection, locality, and freewill loopholes, are interesting open questions for future works. Regardless of these loopholes, the demonstrations with only linear number of measurement settings still could be an efficient fidelity estimation method compared to the traditional tomography method with an exponential number of measurement settings.
	
	Our protocol can be extended to self-test entangled states with more qubits and other types of graph states. After verifying the quantum network with multipartite genuine entanglement, it is also interesting and promising to demonstrate the real application tasks on this distributed quantum network, e.g., measurement-based quantum computation \cite{onewayQC}, secret sharing \cite{markham2008graph, bell2014experimental} under untrusted channels, quantum conference key agreement \cite{murta2020quantum},  and delegated quantum computation \cite{mckague2013interactive, reichardt2013classical}. Moreover, our demonstration and the Bell inequalities constructed from stabilizers might also be extended to detect more detailed entanglement structure \cite{zhou2019detecting}, and self-test high dimensional entangled states and hypergraph states \cite{Rossi_2013}.

	This work was supported by the National Natural Science Foundation of China, the Chinese Academy of Sciences, the National Fundamental Research Program, and the Anhui Initiative in Quantum Information Technologies. Q.~Z.~acknowledges the support by the Department of Defense through the Hartree Postdoctoral Fellowship in the Joint Center for Quantum Information and Computer Science (QuICS).
	
	D.W. and Q.Z. contributed equally to this work.

	Note added.—Recently, we became aware of an independent robust self-testing experiment \cite{xu2021experimental}.

	\bibliographystyle{apsrev4-1}
	
	\bibliography{graphBell}
	
	\onecolumngrid
	\section{supplement information}
	
	\section{1. Theoretical details}
	\subsection{(1). Construction of Bell inequalities for graph states}
	According to the construction procedure in Ref.~\cite{zhao2020constructing}, we first choose a subset of the stabilizer set as $\mathcal{ST}$. 
	We define 
	$\emph{pairable}$ stabilizers in $\mathcal{ST}$  as
	\begin{equation}
	\mathcal{P}=\{(l, k)|S^l,S^k\in \mathcal{ST}~ are~\emph{pairable},~l<k \}.
	\end{equation}
	Here two stabilizers are called \emph{pairable}, if there exists at least one position $i$ such that local operators are anti-commutative.
	We also include the set of all the other non-pairable stabilizers in $\mathcal{ST}$ as $\mathcal{R} $.
	Moreover, we choose some specific positions, denoted as  $\mathcal{AC}$, to replace the operators $X_i$ and $Z_i$ in the stabilizers with the observables $A_i+B_i$ and $A_i-B_i$. 
	
	In a four-party GHZ state, the generators are 
	\begin{equation}
	G_1=X_1Z_2Z_3Z_4,~G_2=Z_1X_2,~G_3=Z_1X_3,~G_4=Z_1X_4.
	\end{equation}
	Here we use stabilizers as 
	\begin{equation}
	\begin{aligned}
	&S^1=G_1=X_1Z_2Z_3Z_4, S^2=G_2=Z_1X_2, S^3=G_3=Z_1X_3,\\ 
	&S^4=G_4=Z_1X_4, S^5=G_2G_3=X_2X_3, S^6= G_2G_4=X_2X_4 .
	\end{aligned}
	\end{equation}
	We construct the following inequalities with various $\mathcal{AC}$, $\mathcal{P}$, $\mathcal{R}$ as the following table. 
	\begin{table}[ht]
		\begin{tabular}{|cl|ccccc|}
			\hline
			&$\mathrm{GHZ}$ & $\mathcal{AC}$ & $\mathcal{P}$ & $\mathcal{R}$  & $\beta_Q$  &$\beta_C$\\[1mm]
			\hline
			& $\mathcal{B}_1 $   &$\{1\}$
			& $\{(1,2)\}$
			& $\{5,6\}$
			& $2\sqrt{2}+2$ & $4$  \\[1mm]
			\hline
			& $\mathcal{B}_2 $ & $\{1\}$
			& $\{(1,2), (1,3)\}$
			& \{6\} & $4\sqrt{2}+1$&$5$  \\[1mm]
			\hline
			& $\mathcal{B}_3 $ & $\{1\}$ & $\{(1,2), (1,3), (1,4)\}$ &  $\varnothing$ & $6\sqrt{2}$ & $6$  \\[1mm]
			\hline
		\end{tabular}
		\caption{4-qubit GHZ state Bell inequalities construction.}
		\label{T:4GHZ}
	\end{table}
	For 4-party Cluster state, we have the generators and use the stabilizers as 
	\begin{equation}
	\begin{aligned}
	&G_1=X_1Z_2,~G_2=Z_1X_2Z_3,~G_3=Z_2X_3Z_4,~G_4=Z_3X_4,\\
	&S^1=G_1=X_1Z_2, S^2=G_2=Z_1X_2Z_3, S^3=G_3=Z_2X_3Z_4, \\
	&S^4=G_4=Z_3X_4, S^5=G_1G_3=X_1X_3Z_4, S^6= G_2G_4=Z_1X_2X_4.
	\end{aligned}
	\end{equation}
	Our constructions are shown in the following table.
	\begin{table}[ht]
		\begin{tabular}{|cl|ccccc|}
			\hline
			\hline
			&$\mathrm{Cluster}$ & $\mathcal{AC}$ & $\mathcal{P}$ & $\mathcal{R}$  &  $\beta_Q$  &$\beta_C$\\[1mm]
			\hline
			& $\mathcal{B}_4 $ & $\{1\}$
			& $\{(1,2)\}$
			& $\{3,4\}$ & $2\sqrt{2}+2$ &4  \\[1mm]
			\hline
			& $\mathcal{B}_5 $& $\{2\}$
			& $\{(1,2), (2,3)\}$
			& \{4\} & $4\sqrt{2}+1$&$5$  \\[1mm]
			\hline
			& $\mathcal{B}_6 $ & $\{2\}$
			& $\{(1,2), (3,6)\}$
			& $\varnothing$ & $4\sqrt{2}$&$4$ \\[1mm]
			
			\hline
		\end{tabular}
		\caption{4-qubit Cluster state Bell inequalities construction. }
		\label{T:cluster}
	\end{table}
	The proof of the classical bounds and quantum bounds are shown in Ref.~\cite{zhao2020constructing}.
	We refer to Ref.~\cite{zhao2020constructing} for more details of construction. 
	
	\subsection{(2). Robust self-test of graph state}
	In this section, we give the detailed explanation about numerical robustness results of self-testing. 
	All the constructed Bell inequalities shown above are suitable for the self-testing of graph states. 
	We could lower bound the fidelity between the measured state $\rho$ and the target graph state $\psi_G$ (under local isometry), with the knowledge of the Bell inequality value. 
	Here one can adopt local extraction channel and the maximal fidelity shows
	\begin{equation}
	\begin{aligned}
	F=\max_{\Lambda=\Lambda_1\otimes\Lambda_2\cdots\Lambda_N } \bra{\psi_G} \Lambda(\rho) \ket{\psi_G},
	\end{aligned}
	\end{equation}
	where $\Lambda_i$ is the local channel on $i$-th party.
	Alternatively, the fidelity can be written as follows
	\begin{equation}
	\begin{aligned}
	F=\Tr[\rho\Lambda_1^\dag \otimes\Lambda_2^\dag \cdots\Lambda_N^\dag (\ket{\psi_G}\bra{\psi_G})],
	\end{aligned}
	\end{equation}
	where $\Lambda_i^\dag$ is the dual channel of $\Lambda_i$.
	Here denote the state after this dual channel as $K=\Lambda_1^\dag \otimes\Lambda_2^\dag \cdots\Lambda_N^\dag (\ket{\psi_G}\bra{\psi_G})$.
	To lower bound $F$ and $K$, we need to choose appropriate parameters $s$ and $\mu$ such that the following inequality on operators always holds
	\begin{equation}\label{Eq:Lbound}
	\begin{aligned}
	K\geq s\mathcal{B}+\mu \mathbb{I},
	\end{aligned}
	\end{equation}
	where $\mathcal{B}$ is Bell inequality to self-test the state. Then the fidelity is bounded as $F\geq s\langle \mathcal{B}\rangle+\mu$, with $\langle \mathcal{B}\rangle=\Tr(\rho\mathcal{B})$ the Bell inequality value. 
	Note that here $\mathcal{B}$ consists of
	the dichotomic  measurements. 
	According to  the Jordan lemma, one can reduce the state to the N-qubit space, and all the possible measurements can be parameterized by  the angles $\theta_i\in[0,\pi/2]$ as
	\begin{equation}
	\begin{aligned}
	A_i&=\cos \theta_i X_i + \sin \theta_i Z_i, B_i=\cos \theta_i X_i - \sin \theta_i Z_i,~i\in \mathcal{AC},\\
	A_i&=\cos \theta_i H_i + \sin \theta_i V_i, B_i=\cos \theta_i H_i - \sin \theta_i V_i, ~i\notin \mathcal{AC},
	\end{aligned}
	\end{equation}
	where  $\mathcal{AC}$ is the rotated set, and  $H_i(V_i)=(X_i\pm Z_i)/\sqrt{2}$.
	Here we consider a specific extraction channel in Ref~\cite{kaniewski2016analytic}
	\begin{equation}
	\begin{aligned}
	\Lambda_i(\rho)=\frac{1+g(x)}{2}\rho+\frac{1-g(x)}{2}\Gamma_i(x)(\rho)\Gamma_i(x),
	\end{aligned}
	\end{equation}
	where $g(x)=(1+\sqrt{2})(\sin x+\cos x+1)$, and  $\Gamma_i(x)$ is the operator on $i$-th qubit: for $i\in T$, $\Gamma_i(x)=X_i(Z_i)$ as $x<  \pi/4$; for $i\notin T$, $\Gamma_i(x)=H_i(V_i)$ as $x<  \pi/4$. 
	Here the Bell inequality $\mathcal{B}$ and the operator $K$ are both parameterized with the parameters $\theta_i$, and the inequality in Eq.~\eqref{Eq:Lbound} is equivalent to the following inequality
	\begin{equation}\label{Eq:Lbound1}
	\begin{aligned}
	K(\theta_1, \theta_2, \cdots, \theta_N) \geq s\mathcal{B}(\theta_1, \theta_2, \cdots, \theta_N)+\mu \mathbb{I}.
	\end{aligned}
	\end{equation}
	There are a lot of suitable $s$ and $\mu$. 
	We should find optimal $s$ and $\mu$ which gives the best lower bound for all possible $\theta_1, \theta_2, \cdots, \theta_N$.
	Here we first fix $s$ and find the minimal eigenvalue of $K(\theta_1, \theta_2, \cdots, \theta_N) \geq s\mathcal{B}(\theta_1, \theta_2, \cdots, \theta_N)$ for all $(\theta_1, \theta_2, \cdots, \theta_N)$. Then we want to find the minimum $s$ satisfying the relation $s\beta_Q+\mu=1$ hold. 
	A smaller $s$ indicates a better bound. Consequently, we list all the obtained $s$ and $\mu$ in Table \ref{Table:smu} as follows.
	\begin{table}[ht]
		\begin{tabular}{|c|cc|c|cc|}
			\hline
			4-qubit GHZ  & $s$ & $\mu$  & 4-qubit Cluster & $s$ & $\mu$ \\[1mm]
			\hline
			$\mathcal{B}_1$ & 1
			&$-1-2\sqrt{2}$& $\mathcal{B}_4$   & 1
			& $-1-2\sqrt{2}$
			\\[1mm]
			\hline
			$\mathcal{B}_2$   & 0.69
			& -3.5931 & $\mathcal{B}_5$   & 0.7400
			& -3.9262
			\\[1mm]
			
			\hline
			$\mathcal{B}_3$   & 0.49
			& -3.1578& $\mathcal{B}_6$    & 0.6200
			& -2.5071
			\\[1mm]
			
			\hline
		\end{tabular}
		\caption{Numerical fidelity bounds for the 4-qubit GHZ state and Cluster state.  }\label{Table:smu}
	\end{table}

	\section{2. Experimental details}
	\subsection{(1). Fidelity estimation}
\begin{figure}[!h]
		\centering
		\begin{minipage}[t]{0.48\textwidth}
			\centering
			\includegraphics[width=9cm]{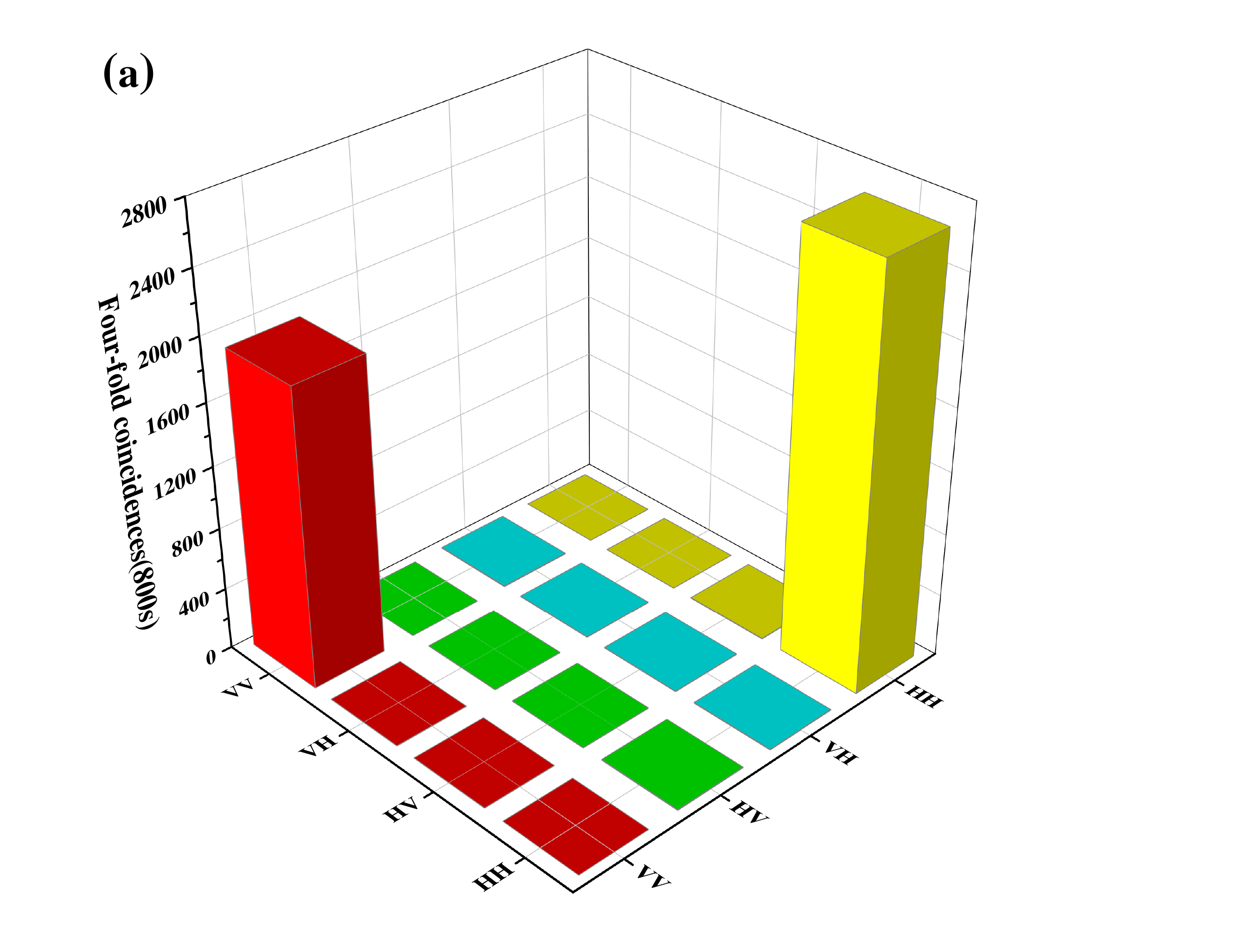}
		\end{minipage}
		\begin{minipage}[t]{0.48\textwidth}
			\centering
			\includegraphics[width=8cm]{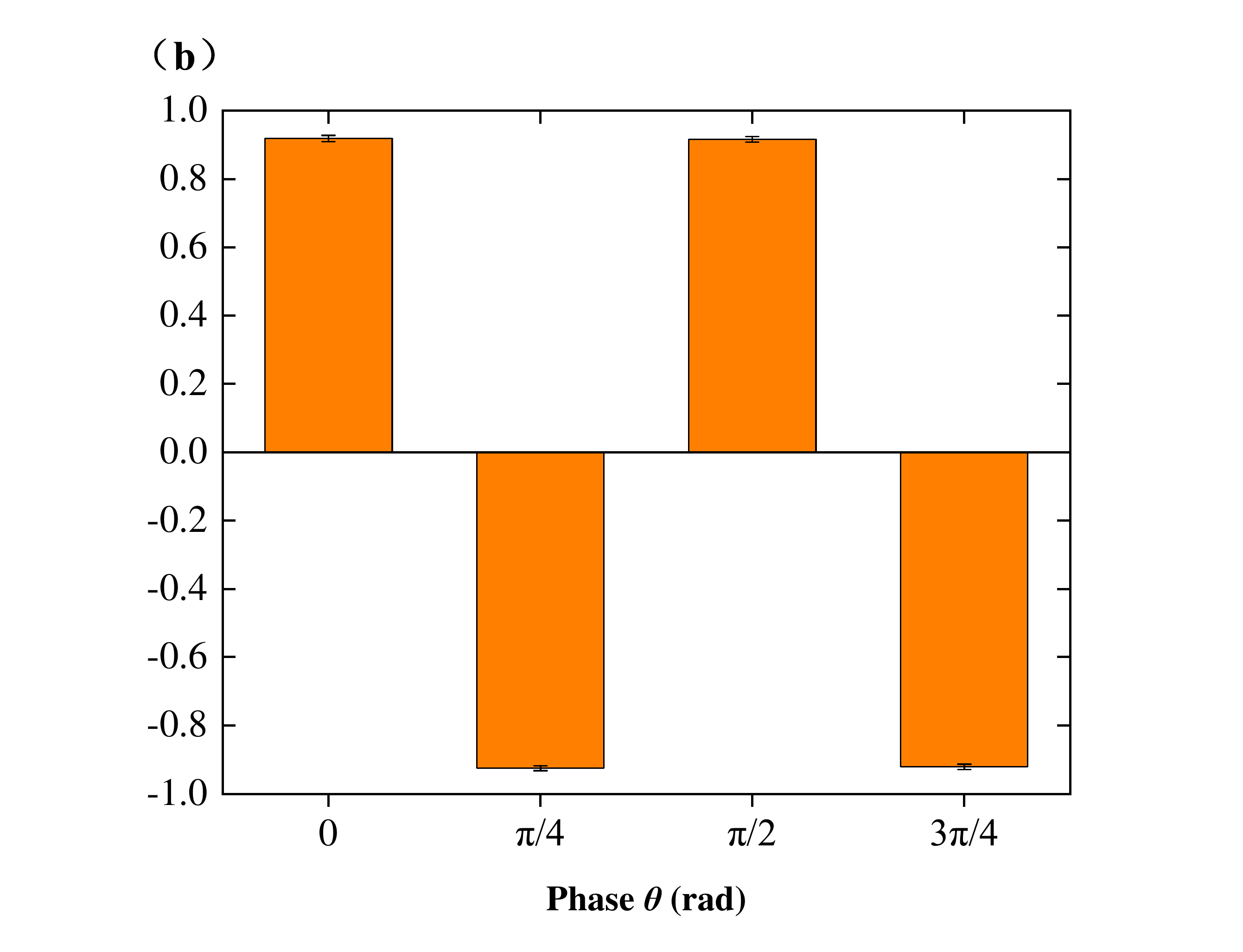}
		\end{minipage}
		\caption{Experimental results of a four-photon GHZ state. (a)
			Fourfold coincidence counts in the H/V basis with data collection in 800s; the value of $(\ket{H}\bra{H})^{\otimes 4}  +  (\ket{V}\bra{V})^{\otimes 4}$ is 0.994(2). (b)
			The expectation values of $ M_{k\pi}^{\otimes 4}= \cos(k\pi/4) \sin( k\pi/4) (k =0, 1, 2, 3)$ are obtained by the measurements in the basis of
			$(\ket{H} \pm e^{ik\pi/4}\ket{V})/\sqrt{2}$. Each basis accumulates about 5000 events
			in 1200s. The error bar is derived from raw detection events
			by Poissonian counting statistics analysis with one standard deviation.
			The measured expectation values are $0.918(9)$, $ -0.924(7)$, $0.916(7)$, and $-0.920(7)$.}
		\label{Fig:GHZstate}
	\end{figure}

	We observe an average two-photon coincidence count rate $\sim$ 23000 Hz with a correlation visibility of $99.6\%$, $99\%$in $\ket{H}/\ket{V}$ and $\ket{+}/\ket{-}$ basis under 450mw laser power, yielding the polarization entanglement of the photon-pair source. For characterization of the N-photon
	states, we measure their state fidelities, that is, the
	overlap of the experimentally produced state with the
	ideal one: $F(\psi)=\bra{\psi^N}\rho_{exp}\ket{\psi^N}=\Tr(\rho_{ideal}\rho_{exp})$, where $\ket{\psi^N}$ denotes the ideal N-photon state and $\rho_{ideal}(\rho_{exp})$ represents the density matrix of the ideal state (experimentally prepared state). For the GHZ state, the fidelity
	can be calculated by the expectation values of the average of $F(GHZ^N)=(\big\langle P^N\big\rangle+\big\langle C^N\big\rangle )/2$,
	with $\big\langle P^N\big\rangle=(\ket{H}\bra{H})^{\otimes N} +  (\ket{V}\bra{V})^{\otimes N}$, which denotes the population of $\ket{H}^{\otimes N}$ and $\ket{V}^{\otimes N}$ are consisted of
	GHZ states. Furthermore, the coherence of the N-qubit GHZ state, which is defined by the off-diagonal element of its density matrix and reflects the coherent superposition between the $\ket{H}^{\otimes N}$ and $\ket{V}^{\otimes N}$component of the GHZ state, can be calculated by
	\begin{equation}
	C^N=\frac{1}{N}\sum_{k=0}^{N-1}(-1)^k\big\langle {M_{(k\pi/N)}}^{\otimes N}\big\rangle,
	\end{equation} 
	with $M_{k\pi/N}=\sigma_xcos(k\pi/N)+\sigma_ysin(k\pi/N)$, where $\sigma_x$ and $\sigma_y $ are Pauli operators and $k=0,1,...,N-1$. We measure the N single photons individually along the basis of $(\ket{H} \pm e^{ik\theta/4}\ket{V})/\sqrt{2}$ and detect all the 2N combinations of N-photon output. From these measurements, we obtain the experimentally estimated expectation values of the observable $ {M_\theta}^{\otimes N}=(cos\theta\sigma_x+sin\theta\sigma_y)$.
	Finally, the population and coherence measured in our experiment is 0.994(2), 0.920(3) are shown in Fig. \ref{Fig:GHZstate}. If two 10nm filters in spatial mode 2 and 3  are replaced with 5nm filters, the fidelity become 0.963(3) , we can get greater violation 4.759(31), 6.544(54), 8.323(77).
	

\begin{figure*}[!h]
		\centering
		\includegraphics[scale=0.4]{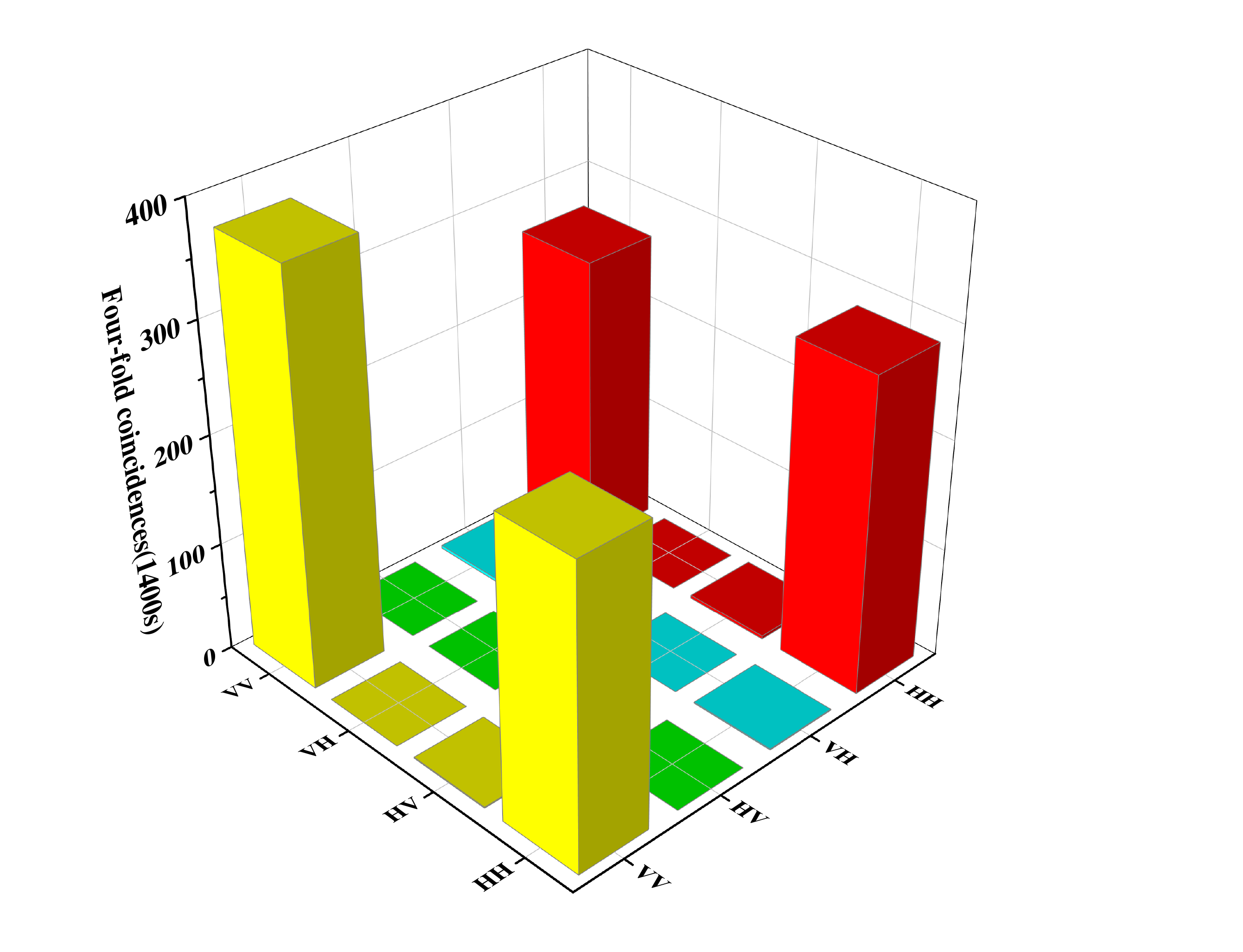}
		\caption{Experiment results of the four-photon cluster state, fourfold coincidence counts acquired in an H/V basis.} 
		\label{Fig:clusterresult}
	\end{figure*}

	To evaluate the four-photon linear cluster state, we estimate its fidelity as $F_{C4}=\bra{C_4}\rho_{exp}\ket{C4}$, by averaging expectation values of sixteen stabilizers .
	As a four-photon linear cluster state, its stabilizer generators are
	\begin{equation}
	g_1=Z_1Z_2I_3I_4, \\
	g_2=X_1X_2Z_3I_4, \\
	g_3=I_1Z_2X_3X_4, \\
	g_4=I_1I_2Z_3Z_4,\\
	\end{equation}
	where $X, Y, Z$, and $I$ denote Pauli operators $\sigma_x$, $\sigma_y$, $\sigma_z$ and the Identity operator, respectively. Therefore, the fidelity for the target state equals to the average expectation value of all the stabilizer operators. The measurement result are shown in Table \ref{Table:stab}. The state fidelity is calculated to be $F_{C4}=0.945(2)$, and the result of fourfold coincidence achieved from $H/V$ basis is shown in Fig. \ref{Fig:clusterresult}.

	\begin{table}[ht]
		\begin{tabular}{ |p{1cm}||p{3cm}|p{3cm}|p{3cm}|  }
			\hline
			& Stabilizer	& Operators  & Expectation value\\
			\hline
			(1)& $g_1$	&$Z_1Z_2I_3I_4$&  0.993(3)	\\
			(2)&	$g_2$	& $X_1X_2Z_3I_4$& 0.930(10)\\
			(3)& $g_3$	& $I_1Z_2X_3X_4$&	0.931(10)\\
			(4)& $g_4$	& $I_1I_2Z_3Z_4$&	0.993(3)\\
			(5)&$g_1g_2$& $-Y_1Y_2Z_3I_4$&0.933(10)\\
			(6)&$g_1g_3$&$Z_1I_2X_3X_4$	&0.927(10)\\
			(7)&$g_1g_4$	& $Z_1Z_2Z_3Z_4$&0.986(4)\\
			(8)&	$g_2g_3$	& $X_1Y_2Y_3X_4$&0.932(10)\\
			(9)& $g_2g_4$	& $X_1X_2I_3Z_4$&0.932(10)\\
			(10)&	$g_3g_4$	& $-I_1Z_2Y_3Y_4$	&0.944(9)\\
			(11)& $g_1g_2g_3$	& $Y_1X_2Y_3X_4$&	0.920(11)\\
			(12)& $g_1g_2g_4$	& $-Y_1Y_2I_3Z_4$&	0.924(11)\\
			(13)&	$g_1g_3g_4$	& $-Z_1I_2Y_3Y_4$&0.942(9)\\
			(14)&	$g_2g_3g_4$	& $X_1Y_2X_3Y_4$	&0.924(11)\\
			(15)&	$g_1g_2g_3g_4$	&$Y_1X_2X_3Y_4$&0.916(11)\\
			(16)&	$I$	& $I_1I_2I_3I_4$&1\\
			\hline
			&$F_{C4}=0.945(2)$&&\\
			\hline
		\end{tabular}\caption{Sixteen expectation values of stabilizers of a four-photon cluster state and their experimental results. The data collection of each is around 1800 events in 1400s and the error bars are deduced from the raw data and Poisson counting statistics.}
		\label{Table:stab}
	\end{table}
	
	\hypertarget{part 02-2}{\subsection{(2). Bell inequalities implementations}}
	In our implementation, the prepared states is close to the targeted graph states up to the local unitary. Thus we have to change the measurement settings accordingly. 
	For GHZ states, the local-unitary is $H_2H_3H_4$ ($H$ is a Hadamard gate, defined as $H =(X+Z)/\sqrt{2}$). Thus the measurements are implemented as $A_1=\frac{X+Z}{\sqrt{2}}, B_1=\frac{X-Z}{\sqrt{2}}$, and $A_i=Z, B_i=X$ when $i \ne 1$. 

		\begin{equation}\label{Eq:GHZ}
		GHZ \left\{
		\begin{aligned}
		&\mathcal{B}_1: \big\langle(A_1+B_1)B_2B_3B_4 \big \rangle+  \big\langle(A_1-B_1)A_2 \big\rangle + \big\langle A_2A_3  \big\rangle+ \big\langle A_2A_4 \big \rangle
		\le \beta_{C,1}=4, \\
		& \mathcal{B}_2 : 2\big\langle(A_1+B_1)B_2B_3B_4 \big \rangle+  \big\langle(A_1-B_1)A_2 \big\rangle +
		\big\langle(A_1-B_1)A_3 \big\rangle+
		\big\langle A_2A_4 \big \rangle
		\le \beta_{C,2}=5, \\
		&   \mathcal{B}_3: 3\big\langle(A_1+B_1)B_2B_3B_4 \big \rangle+  \big\langle(A_1-B_1)A_2 \big\rangle +
		\big\langle(A_1-B_1)A_3 \big\rangle+
		\big\langle(A_1-B_1)A_4 \big\rangle
		\le \beta_{C,3}=6 .\\
		\end{aligned}
		\right.
		\end{equation}

	For cluster states, the local-unitary is $ H_1H_4$. For the Bell inequality $\mathcal{B}_4$, we  set
	$A_1=\frac{X+Z}{\sqrt{2}}, B_1=\frac{Z-X}{\sqrt{2}}$, $A_i=X, B_i=Z$ $(i=2,3)$, $A_4=Z$, $B_4=X$, and $A_i=Z, B_i=X (i=1,4)$, $A_2=\frac{X+Z}{\sqrt{2}}, B_2=\frac{X-Z}{\sqrt{2}}$, $A_3=X, B_3=Z$ for $\mathcal{B}_5$ and $\mathcal{B}_6$. 

		\begin{equation}
		Cluster \left\{
		\begin{aligned}
		&\mathcal{B}_4: \big  \langle(A_1+B_1)B_2\big  \rangle+ \big  \langle(A_1-B_1)A_2B_3  \big  \rangle +\big  \langle B_2A_3B_4 \big \rangle+\big  \langle B_3A_4 \big  \rangle \le \beta_{C,1}=4,\\
		&  \mathcal{B}_5: \big  \langle A_1(A_2-B_2)\big  \rangle+ 2\big  \langle B_1(A_2+B_2)B_3  \big  \rangle +\big  \langle (A_2-B_2)A_3B_4 \big \rangle+\big  \langle B_3A_4 \big  \rangle \le \beta_{C,2}=5,\\
		& \mathcal{B}_6: \big  \langle A_1(A_2-B_2)\big  \rangle+ \big  \langle B_1(A_2+B_2)B_3  \big  \rangle +\big  \langle (A_2-B_2)A_3B_4 \big \rangle+\big  \langle B_1(A_2+B_2)A_4 \big  \rangle \le \beta_{C,3}=4.
		\\
		\end{aligned}
		\right.
		\end{equation}

	We show the experimental results for different measurement settings and and the corresponding Bell values with various degree of noise in Tables.\ref{table:mytable1}-\ref{table:mytable3} as well as the comparison with the entanglement witness in Fig.\ref{Fig:result}.
	
	
	\begin{table}[ht]
		\renewcommand{\arraystretch}{1.4}
		\begin{tabular}{|>{\centering}m{5.em}|c|>{\centering}m{5.em}|c|>{\centering}m{5.em}|c|>{\centering}m{4.em}|c|>{\centering}m{4.em}|c|>{\centering}m{4.em}|c|}
			\hline
			\multicolumn{6}{|c|}{GHZ}& \multicolumn{6}{c|}{Cluster}\\ \hline
			\multicolumn{2}{|c|}{$\mathcal{B}_1$}& \multicolumn{2}{c|}{$\mathcal{B}_2$} & \multicolumn{2}{c|}{$\mathcal{B}_3$} & \multicolumn{2}{c|}{$\mathcal{B}_4$} & \multicolumn{2}{c|}{$\mathcal{B}_5$} & \multicolumn{2}{c|}{$\mathcal{B}_6$} \\ \hline
			B2B3 & 0.992(2) & A1B3 & 0.738(10) & A1B3 & 0.738(10)  & A1B2 & 0.739(20) & B1A2 & 0.655(23)  & B1A2 & 0.655 (23)\\
			
			B2B4 & 0.991(2)  & B2B4 & 0.991(2) & A1B2 & 0.739(10)  & B1B2 & 0.668(22) & B1B2 & 0.779(19) & B1B2 & 0.779(19)\ \\
			
			A1B2 & 0.739(10)  & A1B2 & 0.739(10)  & B1B2 & 0.681(11) & A1A2B3 & 0.610(24) & A1A2B3 & 0.695(21) & A1A2B3 & 0.695(21)\ \\
			
			B1B2 & 0.681(11)  & B1B2& 0.681(11) & A1A2A3A4 & 0.656(11)  & B1A2B3 & 0.727(20) & A1B2B3 & 0.642(24) & A1B2B3 & 0.642(24)\  \\
			
			A1A2A3A4 & 0.656(11)  & A1A2A3A4 &0.656(11)   & B1A2A3A4 & 0.680(11)  & B2A3A4 & 0.931(10)  & A2A3A4 &0.642(22) & A2A3A4 &0.653(22)\   \\
			
			B1A2A3A4 &0.680(11)  &  B1A2A3A4 & 0.680(11)   & B1B3 & 0.681(11)  & B3B4 & 0.993(3)  & B2A3A4 & 0.689(22) & B2A3A4 & 0.689(22)\ \\
			
			&   &  B1B3 & 0.681(11) & A1B4 &0.738(10) &   &   & B3B4 & 0.993(3) & A1A2B4 & 0.685(22)\ \\
			
			&   &   &   &  B1B4 & 0.682(11)  &   &   &  &   & A1B2B4 & 0.633(24)\\\hline
			\multicolumn{2}{|c|}{\big\langle$\mathcal{B}_1$\big\rangle=4.738(21)}& \multicolumn{2}{c|}{\big\langle$\mathcal{B}_2$\big\rangle=6.501(37)} & \multicolumn{2}{c|}{\big\langle$\mathcal{B}_3$\big\rangle=8.266(53)} & \multicolumn{2}{c|}{\big\langle$\mathcal{B}_4$\big\rangle=4.669(45)} & \multicolumn{2}{c|}{\big\langle$\mathcal{B}_5$\big\rangle=6.434(77)} & \multicolumn{2}{c|}{\big\langle$\mathcal{B}_6$\big\rangle=5.431(62)} \\ \hline 
			
		\end{tabular}\caption{Bell values of every measurement setting in the experiment}
		\label{table:mytable1}
	\end{table}
	
	\begin{table}[ht]
		\renewcommand{\arraystretch}{1.4}
		\begin{tabular}{|>{\centering}m{5.em}|c|>{\centering}m{5.em}|c|>{\centering}m{5.em}|c|>{\centering}m{4.em}|c|>{\centering}m{4.em}|c|>{\centering}m{4.em}|c|}
			\hline
			\multicolumn{6}{|c|}{GHZ}& \multicolumn{6}{c|}{Cluster}\\ \hline
			\multicolumn{2}{|c|}{$\mathcal{B}_1$}& \multicolumn{2}{c|}{$\mathcal{B}_2$} & \multicolumn{2}{c|}{$\mathcal{B}_3$} & \multicolumn{2}{c|}{$\mathcal{B}_4$} & \multicolumn{2}{c|}{$\mathcal{B}_5$} & \multicolumn{2}{c|}{$\mathcal{B}_6$} \\ \hline
			B2B3 & 0.992(2) & A1B3 & 0.740(9) & A1B3 & 0.740(9)  & A1B2 & 0.731(20) & B1A2 & 0.659(22)  & B1A2 & 0.659 (22)\\
			
			B2B4 & 0.989(2)  & B2B4 & 0.989(2) & A1B2 & 0.739(9)  & B1B2 & 0.663(21) & B1B2 & 0.769(18) & B1B2 & 0.769(18)\ \\
			
			A1B2 & 0.739(9)  & A1B2 & 0.739(9)  & B1B2 & 0.643(12) & A1A2B3 & 0.666(21) & A1A2B3 & 0.633(22) & A1A2B3 & 0.633(22)\ \\
			
			B1B2 & 0.643(11)  & B1B2& 0.643(12) & A1A2A3A4 & 0.597(11)  & B1A2B3 & 0.845(15) & A1B2B3 & 0.590(24) & A1B2B3 & 0.590(24)\  \\
			
			A1A2A3A4 & 0.597(11)  & A1A2A3A4 &0.597(11)   & B1A2A3A4 & 0.619(11)  & B2A3A4 & 0.994(3)  & A2A3A4 &0.585(22) & A2A3A4 &0.592(22)\   \\
			
			B1A2A3A4 &0.619(11)  &  B1A2A3A4 & 0.619(11)   & B1B3 & 0.659(11)  & B3B4 & 0.991(2)  & B2A3A4 & 0.621(22) & B2A3A4 & 0.621(22)\ \\
			
			&   &  B1B3 & 0.659(11) & A1B4 &0.739(9) &   &   & B3B4 & 0.994(3) & A1A2B4 & 0.625(22)\ \\
			
			&   &   &   &  B1B4 & 0.654(11)  &   &   &  &   & A1B2B4 & 0.577(24)\\\hline
			\multicolumn{2}{|c|}{\big\langle$\mathcal{B}_1$\big\rangle=4.579(21)}& \multicolumn{2}{c|}{\big\langle$\mathcal{B}_2$\big\rangle=6.202(38)} & \multicolumn{2}{c|}{\big\langle$\mathcal{B}_3$\big\rangle=7.821(54)} & \multicolumn{2}{c|}{\big\langle$\mathcal{B}_4$\big\rangle=4.459(46)} & \multicolumn{2}{c|}{\big\langle$\mathcal{B}_5$\big\rangle=6.072(77)} & \multicolumn{2}{c|}{\big\langle$\mathcal{B}_6$\big\rangle=5.064(62)} \\ \hline 
			
		\end{tabular}\caption{Noise values of every measurement setting in the experiment when $p=0.1$}
		\label{table:mytable2}
	\end{table}

	\begin{table}[ht]
		\renewcommand{\arraystretch}{1.4}
		\begin{tabular}{|>{\centering}m{5.em}|c|>{\centering}m{5.em}|c|>{\centering}m{5.em}|c|>{\centering}m{4.em}|c|>{\centering}m{4.em}|c|>{\centering}m{4.em}|c|}
			\hline
			\multicolumn{6}{|c|}{GHZ}& \multicolumn{6}{c|}{Cluster}\\ \hline
			\multicolumn{2}{|c|}{$\mathcal{B}_1$}& \multicolumn{2}{c|}{$\mathcal{B}_2$} & \multicolumn{2}{c|}{$\mathcal{B}_3$} & \multicolumn{2}{c|}{$\mathcal{B}_4$} & \multicolumn{2}{c|}{$\mathcal{B}_5$} & \multicolumn{2}{c|}{$\mathcal{B}_6$} \\ \hline
			B2B3 & 0.993(2) & A1B3 & 0.736(9) & A1B3 & 0.736(9)  & A1B2 & 0.730(19) & B1A2 & 0.666(20)  & B1A2 & 0.666 (20)\\
			
			B2B4 & 0.989(2)  & B2B4 & 0.988(2) & A1B2 & 0.740(9)  & B1B2 & 0.656(20) & B1B2 & 0.764(18) & B1B2 & 0.764(18)\ \\
			
			A1B2 & 0.740(9)  & A1B2 & 0.740(9)  & B1B2 & 0.620(11) & A1A2B3 & 0.516(24) & A1A2B3 & 0.568(22) & A1A2B3 & 0.568(22)\ \\
			
			B1B2 & 0.620(11)  & B1B2& 0.620(11) & A1A2A3A4 & 0.547(11)  & B1A2B3 & 0.607(22) & A1B2B3 & 0.546(24) & A1B2B3 & 0.546(24)\  \\
			
			A1A2A3A4 & 0.547(11)  & A1A2A3A4 &0.547(11)   & B1A2A3A4 & 0.564(11)  & B2A3A4 & 0.784(16)  & A2A3A4 &0.529(22) & A2A3A4 &0.529(22)\   \\
			
			B1A2A3A4 &0.564(11)  &  B1A2A3A4 & 0.564(11)   & B1B3 & 0.632(10)  & B3B4 & 0.994(3)  & B2A3A4 & 0.578(22) & B2A3A4 & 0.578(22)\ \\
			
			&   &  B1B3 & 0.632(10) & A1B4 &0.741(9) &   &   & B3B4 & 0.994(3) & A1A2B4 & 0.588(22)\ \\
			
			&   &   &   &  B1B4 & 0.631(10)  &   &   &  &   & A1B2B4 & 0.531(24)\\\hline
			\multicolumn{2}{|c|}{\big\langle$\mathcal{B}_1$\big\rangle=4.452(21)}& \multicolumn{2}{c|}{\big\langle$\mathcal{B}_2$\big\rangle=5.940(37)} & \multicolumn{2}{c|}{\big\langle$\mathcal{B}_3$\big\rangle=7.435(53)} & \multicolumn{2}{c|}{\big\langle$\mathcal{B}_4$\big\rangle=4.288(45)} & \multicolumn{2}{c|}{\big\langle$\mathcal{B}_5$\big\rangle=5.758(77)} & \multicolumn{2}{c|}{\big\langle$\mathcal{B}_6$\big\rangle=4.769(62)} \\ \hline 
			
		\end{tabular}\caption{Noise values of every measurement setting in the experiment when $p=0.2$}
		\label{table:mytable3}
	\end{table}
	
	\begin{figure*}[ht]
		\centering
		\includegraphics[width=1\textwidth]{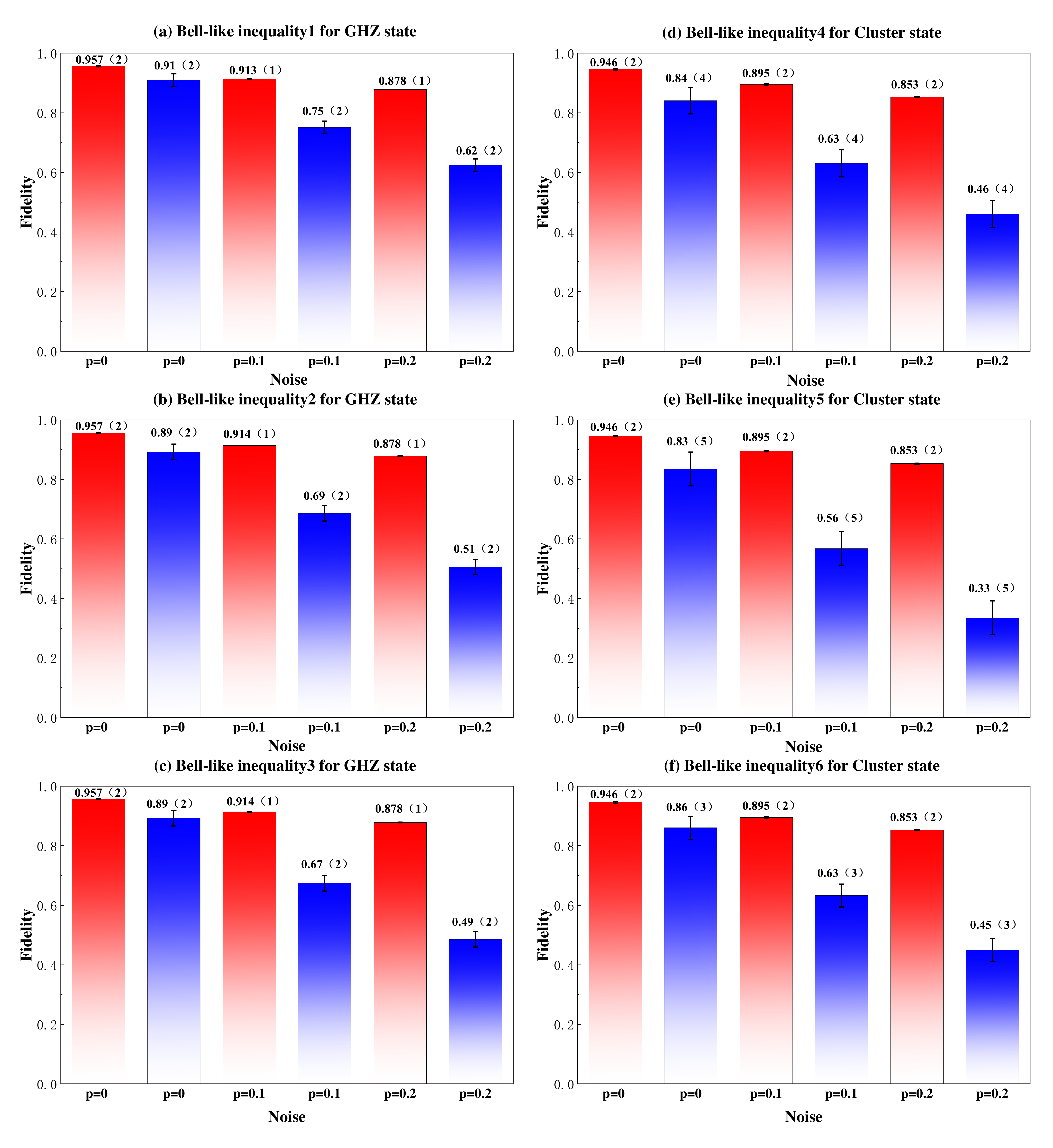}
		\caption{Red columns show the device dependent fidelity estimation obtained from entanglement witness. Blue columns show the device independent fidelity estimation via the self-testing. $p=0, 0.1, 0.2$ means the fraction of the noise state preparing in the experiment.   } 
		\label{Fig:result}
	\end{figure*} 


\end{document}